\documentclass[useAMS,usenatbib]{mn2e}
\usepackage{graphicx}

\newcommand{\pq}{\ensuremath{P_Q}}
\newcommand{\pu}{\ensuremath{P_U}}

\newcommand{\nnq}{\ensuremath{N_Q}}
\newcommand{\nnu}{\ensuremath{N_U}}

\newcommand{\ainv}{\ensuremath{\alpha_{\rm inv}}}
\newcommand{\amin}{\ensuremath{\alpha_{\rm min}}}
\newcommand{\pmin}{\ensuremath{P_Q^{\rm (min)}}}

\title[Polarimetry of (101955) Bennu]{Unusual polarimetric properties of (101955) Bennu:
similarities with F-class asteroids and cometary bodies}
\author[]{A. Cellino$^{1}$, S. Bagnulo$^{2}$, I. N. Belskaya$^{3}$, A.A. Christou$^{2}$\\
$^{1}$INAF - Osservatorio Astrofisico di Torino, I-10025 Pino Torinese, Italy.
{\rm E-mail: alberto.cellino@inaf.it}\\
$^{2}$Armagh Observatory and Planetarium, College Hill, Armagh BT61 9DG, UK.
{\rm E-mail: stefano.bagnulo@armagh.ac.uk}\\
$^{3}$Institute of Astronomy of V.N. Karazin Kharkiv National University, Sumska Str. 35, 
Kharkiv 61022, Ukraine.
{\rm E-mail: irina@astron.kharkov.ua}\\
}

\begin{document}

\date{Accepted 2018 August 23. Received 2018 August 10; in original form 2018 July 15}

\pagerange{\pageref{firstpage}--\pageref{lastpage}} \pubyear{2018}

\maketitle

\label{firstpage}

\begin{abstract}
  We have obtained polarimetric measurements of asteroid (101955)
  Bennu, a presumably primitive near-Earth object (NEO) that is the
  target of NASA's sample return mission OSIRIS-REx. During our
  observing campaign, Bennu was visible from Earth under a wide range
  of illumination conditions, with phase-angle in the range
  16\degr\ to 57\degr. Together with (3200) Phaethon and (152679) 1998 KU2, 
  observed very recently, Bennu is
  the only existing example of a primitive NEO observed in
  polarimetric mode over a wide interval of phase angles. Based on our
  polarimetric data, we propose that Bennu belongs to the unusual
  F taxonomic class defined in the 80s. According to previous works,
  the F-class includes objects with cometary features. This fact can
  be of great importance for the interpretation of the results of the
  exploration of this object by OSIRIS-REx. From polarimetry we 
  also derive an estimate of the geometric albedo of Bennu:
  $p_R=0.059 \pm 0.003$.
\end{abstract}


\section{Introduction}\label{Sect_Intro}

Asteroid (101955) Bennu is the asteroid target of the OSIRIS-REx space
mission \citep{OSIRIS-REx}. From ground-based observations we know
that this object is 550 meters in size and has a rather spheroidal
shape \citep[][and references therein]{Laurettaetal15}.  By applying
to radar and thermal IR observations a sophisticated thermophysical
model, \citet{Yu&Ji15} found for Bennu a geometric albedo of
$0.047^{+0.008}_{-0.001}$, and a thermal inertia suggesting a surface
covered by a fine-grained regolith.  Its spectral reflectance
properties make Bennu a member of the B taxonomic class, according to
the SMASS-based classification by \citet{BusBin02}.  Bennu was chosen
as the target of OSIRIS-REx because it satisfied some basic
requirements: it is representative of the population of primitive,
low-albedo asteroids orbiting in the inner Solar System, which are
thought to be the parent bodies of the most ancient classes of
primitive meteorites, including Carbonaceous Chondrites. Moreover, it
has a very suitable orbit for a sample-return mission.

The modern B taxonomic class as defined by \citet{BusBin02} (SMASS taxonomy)
includes asteroids that in the 80s were classified into two separate classes,
named B and F. These classes were distinguished on the basis of subtle
differences in the spectral reflectance behaviour at $\lambda \le
0.4\,\mu$m. In particular, the F class, first proposed by
\citet{GraTed82}, exhibits a flat spectrophotometric trend over the
whole interval of covered wavelengths, including the blue region,
whereas other classes of objects exhibiting an overall flat spectral
trend show a clear decrease of flux shortward of 0.4\,$\mu$m
\citep[see Fig.~9 of][]{Tholen84}. Because the bluest spectral region
is rarely observed in modern CCD-based spectroscopic surveys,
asteroids originally classified as F belong now to the modern B class
defined by \citet{BusBin02}, characterized by a generally flat or
slightly blueish reflectance spectrum over an interval of wavelengths
between 0.5 and 1\,$\mu$m.  For example, asteroid (2) Pallas is the
largest member of the B class in both the SMASS and in the Tholen's
taxonomy, whereas the largest F-class asteroid, (704) Interamnia, is B-type in
the current SMASS taxonomy.

Apart from their spectrophotometric properties, asteroids can be
distinguished also by measuring the varying state of polarisation of
the sunlight scattered by their surfaces in different illumination
conditions \citep[see, e.g.][]{Kolobook}.  The main source of
information are the so-called phase-polarisation curves, i.e, the
polarisation value as a function of the phase-angle (which is the
angle between the Sun, the target, and the observer). Asteroids with
similar spectroscopic characteristics often share similar polarimetric
properties \citep{Penetal05,Beletal17}. There are, moreover, cases in
which the polarimetric behaviour sharply characterizes some objects
that would be ambiguously characterized based on reflectance spectra.
The two most important cases are those of the so-called Barbarians
\citep{Celetal06}, which are beyond the scope of the present paper,
and of the asteroids previously classified as members of the old F
taxonomic class \citep{Beletal05}. In both cases, we find
unusual values of the so-called inversion angle of polarisation.

It is known that at small-phase angles, atmosphere-less objects of our
solar system exhibit the phenomenon of so-called ``negative''
polarisation, which means that the scattered sunlight is linearly
polarised in the direction parallel to the scattering plane.  For the
vast majority of asteroids the polarisation changes its sign (i.e.,
becomes perpendicular to the scattering plane) at phase-angles $\ga
20\degr$ (the "inversion angle" \ainv). However, in the case of the
F-class asteroids, the inversion angle occurs at the distinctly lower
phase-angle of $\sim 16\degr$ \citep{Beletal17}. Furthermore, the
slope of the polarimetric curve around the inversion angle tends to be
substantially steeper for F-class asteroids with respect to other
taxonomic classes. It is not yet fully understood what physical properties 
are responsible for such behaviour, but it is believed that it is due to 
an interplay of composition, refractive index and sizes of surface regolith 
particles \citep{Beletal05}.

Although no longer retained in modern classification systems, the old
distinction between F-class and B-class asteroids is therefore
meaningful, and is very interesting also in other
respects. \citet{Celetal01} noted that the Polana family, located
in a region of the asteroid main belt which may be an important source of NEOs,   
includes several F-class members \citep{Botetal15,deLetal18}. 
Perhaps more interestingly, the peculiar polarimetric properties of the F-class
asteroids, in particular a low value of the polarimetric inversion angle, 
seem to be shared also by cometari nuclei, in a few cases of polarimetric 
measurements of cometary nuclei observed in conditions of absence 
of the coma \citep{Stinson17}.  In the past, the object (4015) Wilson-Harrington,
originally discovered and classified as an F-class asteroid in 1992, was later 
found to exhibit cometary activity \citep[see][and references
  therein]{Fernandezetal97}. 133P/Elst-Pizarro, an object described as
either an active asteroid or a cometary object in the main asteroid
belt, which exhibits recurrent cometary activity, was also found by
\citet{Bagetal10} to exhibit a low value of the inversion angle of polarisation. There is therefore some evidence
that at least a subset of the old F-class asteroids may include
active, or sporadic, or nearly extinct (as in the case of
Wilson-Harrington) comets \citep[see also][]{BelAstIV, Kolobook}.

Knowing that it was classified as B-class, we have decided to carry out
a polarimetric investigation of (101955) Bennu, and perform a
comparison with the polarimetric properties of F-type objects.

\section{Observations}
\begin{table*}
\caption{\label{Tab_Log}
Polarimetry of asteroid (101955) Bennu in the FORS2 $R$ special filter.
$\pq$ and $\pu$ are the reduced Stokes parameters measured in a reference
system such that $\pq$ is the flux perpendicular to the plane Sun-Object-Earth (the
scattering plane) minus the flux parallel to that plane, divided by
the sum of the two fluxes.
}
\begin{center} 
\begin{tabular}{ccrcr@{\,$\pm$\,}lr@{\,$\pm$\,}lr@{\,$\pm$\,}lr@{\,$\pm$\,}l}
\hline \hline
Date                             & 
Time (UT)                        & 
\multicolumn{1}{c}{Exp}          & 
Phase angle                      & 
\multicolumn{2}{c}{\pq}          & 
\multicolumn{2}{c}{\pu}          & 
\multicolumn{2}{c}{\nnq}         & 
\multicolumn{2}{c}{\nnu}        \\ 
\multicolumn{1}{c}{(yyyy mm dd)} &  
\multicolumn{1}{c}{(hh:mm)}      &  
(sec)                            &  
\multicolumn{1}{c}{(DEG)}        &  
\multicolumn{2}{c}{(\%)}         &  
\multicolumn{2}{c}{(\%)}         &  
\multicolumn{2}{c}{(\%)}         &  
\multicolumn{2}{c}{(\%)}        \\  
\hline
            &       &       &       & \multicolumn{5}{c}{}                                    \\
 2018 02 19 & 08:37 & 2760  & 57.12 & $ 21.05$& 0.56 &\multicolumn{2}{c}{---} &$-0.48$&0.56 &\multicolumn{2}{c}{---} \\ 
 2018 02 20 & 08:09 & 5520  & 56.90 & $ 20.78$& 0.48 &\multicolumn{2}{c}{---} &$-0.46$&0.48 &\multicolumn{2}{c}{---} \\ 
 2018 03 18 & 06:39 & 4400  & 49.56 & $ 15.51$& 0.33 &$ -0.36    $&   0.30    &$ 0.46$&0.33 &$-0.11$&   0.30    \\ 
 2018 03 21 & 06:52 & 4400  & 48.43 & $ 15.34$& 0.45 &$ -0.28    $&   0.50    &$-0.22$&0.45 &$ 0.73$&   0.50    \\ 
 2018 05 12 & 03:31 & 3520  & 15.96 & $ -0.52$& 0.12 &$ -0.17    $&   0.12    &$ 0.30$&0.12 &$ 0.14$&   0.12    \\ 
 2018 06 11 & 01:58 & 3520  & 27.89 & $  3.95$& 0.16 &$  0.14    $&   0.13    &$-0.02$&0.16 &$-0.15$&   0.13    \\ 
 2018 06 16 & 00:37 & 3520  & 31.24 & $  5.19$& 0.14 &$ -0.44    $&   0.14    &$-0.14$&0.14 &$-0.16$&   0.14    \\ 
\hline
\end{tabular}
\end{center}

\noindent
\begin{small}
  
\end{small}
\end{table*}

\begin{figure*}
  \includegraphics*[angle=270,width=15.5cm,trim={1.0cm 2.0cm 1.8cm 0.6cm},clip]{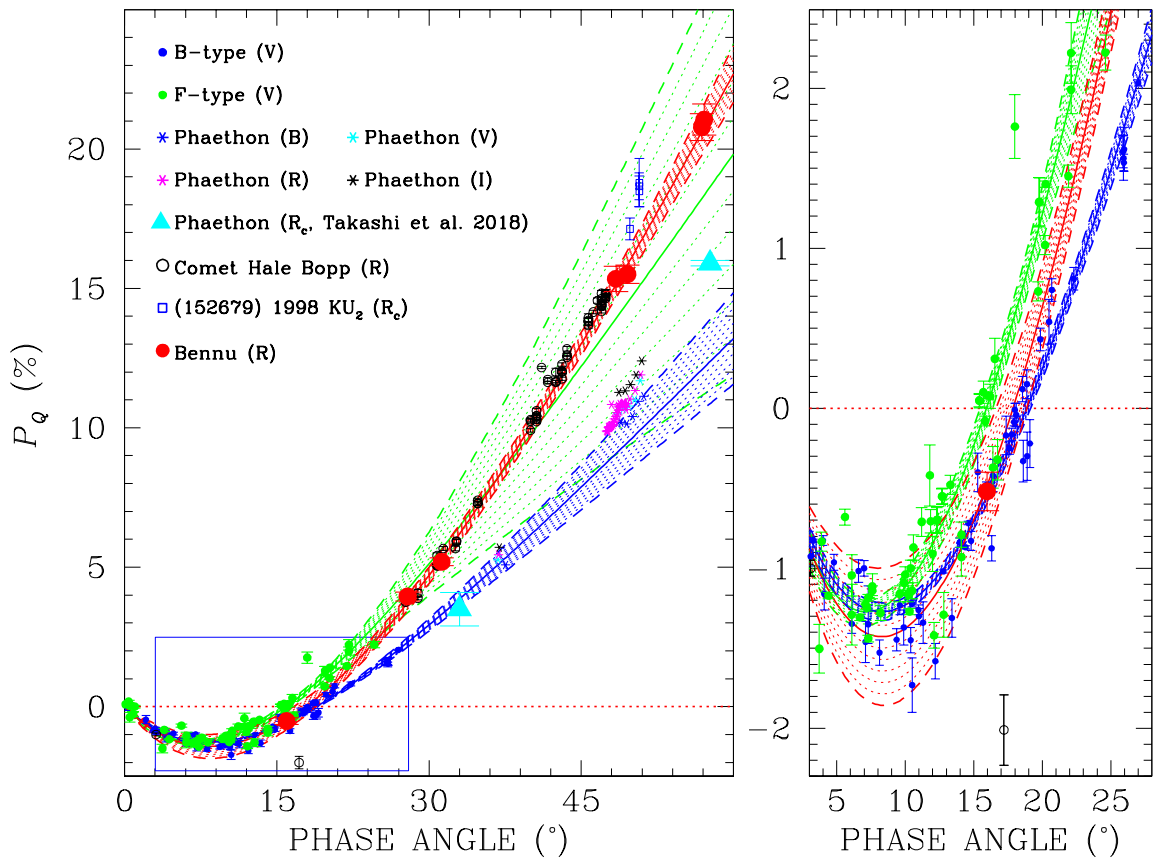}
\caption{\label{Fig_PQ} Polarisation
  vs.\ phase-angle for asteroid Bennu compared with B-class and F-class main belt
  asteroids, near-Earth asteroids (3200) Phaethon and (152679) 1998 KU$_2$, 
  and comet Hale-Bopp. Key to the symbols
  is given in the left panel. Solid lines and dotted areas delimited by dashed lines
  show the best-fits and $3\,\sigma$ boundaries 
  to data for
  Bennu (red lines), B (blue lines), and F (green lines) asteroids.
}
\end{figure*}
Broadband linear polarisation measurements of (101955) Bennu were
obtained using the FORS2 instrument \citep{Appetal98} of the ESO Very
Large Telescope (VLT) in the $R$ special filter 
(centered at 0.655 $\mu$m, $FWHM = 0.165$ $\mu$m),
which was chosen to
maximise the {\it S/N} ratio. 
During ESO period P100 (in February and
March 2018) we obtained three FORS2 polarimetric measurements at large
phase-angles. This dataset was then complemented with three additional
measurements obtained in May and June 2018 thanks to Director
Discretionary Time. Observations were acquired using the beam-swapping
technique \citep[see, e.g.,][]{Bagetal09}, which consists of obtaining
flux measurements at different position angles of the $\lambda/2$
retarder waveplate: 0\degr, 22.5\degr, \ldots, 157.5\degr.  Data
reduction was performed as
described in detail
in Sect.~2.2 of \citet{Bagetal16}.
Throughout this paper we will refer to the reduced Stokes parameter
$\pq = Q/I$ representing the flux perpendicular to the plane
Sun-Object-Earth (the scattering plane) minus the flux parallel to
that plane, divided by the sum of the two fluxes. For symmetry
reasons, the reduced Stokes parameter $\pu = U/I$ is expected to be
zero. In fact, in two observing epochs (19 and 20 February 2018)
the instrument position angle
was set parallel to the scattering plane, and only \pq\ was
measured, under the assumption that the polarisation was either
perpendicular or parallel to the scattering plane (this condition is
nearly always satisfied for all asteroids at all phase-angles). As an
additional quality check, we provided null parameters \nnq\ and \nnu,
which are expected to be consistent with zero within error bars
\citep{Bagetal09}.

The observing log and our measurements are given in
Table~\ref{Tab_Log}, and plotted in Fig.~\ref{Fig_PQ},
together with literature measurements of other objects as discussed in
Sect.~\ref{Sect_Analysis}.

\section{Analysis}\label{Sect_Analysis} 
A comparison of our polarimetric data of Bennu with literature data of
other asteroids is not straightforward because most polarimetric
measurements have been obtained for main-belt asteroids that cannot be
observed at phase-angles $\ga 30^\circ$. At these phase angles,
differences in the polarimetric behaviour of objects of different
classes may be firmly detected only when several data points measured
with accuracy better than $\sim 0.1$\,\% are available. Asteroid Bennu
is faint ($V\ga 20$), and even with a 8\,m class telescope it was not
possible to densely monitor its polarimetric behaviour with
phase-angle with uncertainties smaller than 0.1--0.15\,\%.  In the
positive branch, linear polarization increases significantly with
phase-angle, and at $\alpha \ga 30\degr$, corresponding to most of our
observations of Bennu, the differences between the polarimetric curves
of asteroids of differents spectral classes may be detected with lower
{\it S/N} measurements, but literature data are scarce to make a
statistically meaningful comparison with other known objects.

We should also remind the reader that most polarimetric observations of
asteroids were obtained in the $V$ filter, while Bennu was observed in
the $R$ special filter. Therefore some caution is needed to make a
comparison between our data and those from previous
literature. Spectropolarimetry of asteroid (2) Pallas by
\citet{Bagetal15} shows that in the positive branch, both at
$\alpha=22.9\degr$ and $\alpha=27.5\degr$, polarisation is nearly
constant with wavelength in the optical up to 900\,nm. A similar
property was also found by \citet{Devetal18} for the B-type asteroid
(3200) Phaethon at phase-angles similar to those of our observations
of Bennu. Furthermore, \citet{Giletal14} presented thirty-eight
polarimetric observations of seven asteroids belonging to the
SMASS-based B-class, and they merged together $V$ and $R$ data in
building averaged phase-polarisation curves for these objects, since
no systematic difference was found for data taken in the two different
colours. We conclude that our polarimetric observations in the $R$
band may be compared meaningfully with those of literature data of
F-type and B-type asteroids obtained both in $R$ and in $V$ filters.

In our analysis, we have fit our polarimetric data of Bennu with
the linear-exponential function
\begin{equation}
  \pq(\alpha) = A \left({\rm e}^{-(\alpha/B)} - 1\right) + C\,\alpha
\label{Eq_Fit}
\end{equation}
This is a three-parameter relation issued from a
semi-empirical modeling that \citet{Shennaetal2003} found to be suited to 
fit both phase-magnitude relations in asteroid photometry,
and phase-polarization curves in asteroid polarimetry
\citep[see also][and references therein]{Beletal17, Celetal15}. In Eq.~(\ref{Eq_Fit}), 
$\alpha$ is the phase-angle, and $A$, $B$ and $C$ are free parameters.  
The red solid-line in Fig.~\ref{Fig_PQ} shows our best-fit obtained with
Eq.~(\ref{Eq_Fit}) together with $\pm 3\,\sigma$ uncertainties
represented by red dashed-lines.

\begin{table}
  \caption{\label{Tab_Best-Fit} Best-fit parameters}
  \begin{center}
  \begin{tabular}{lr@{$\pm$}l r@{$\pm$}l r@{$\pm$}l r@{$\pm$}l}
    \hline\hline
    OBJECT&
    \multicolumn{2}{c}{\ainv}         &
    \multicolumn{2}{c}{\amin}        &
    \multicolumn{2}{c}{\pmin}           &
    \multicolumn{2}{c}{Slope}      \\
                                        &
    \multicolumn{2}{c}{(\degr)}         &
    \multicolumn{2}{c}{(\degr)}        &
    \multicolumn{2}{c}{(\%)}           &
    \multicolumn{2}{c}{(\%/\degr)}      \\
    \hline
Bennu   & 17.88& 0.40 & 8.28 & 0.13 & $-1.43$ & 0.14 & 0.276 &0.012 \\
B class & 18.86& 0.09 & 8.26 & 0.10 & $-1.27$ & 0.02 & 0.210 &0.003 \\
F class & 15.93& 0.15 & 7.15 & 0.22 & $-1.27$ & 0.03 & 0.260 &0.008 \\
\hline
  \end{tabular}
  \end{center}
  \end{table}

Using Eq.~(\ref{Eq_Fit}) we have built two synthetic
phase-polarisation curves by merging literature data of asteroids (2),
(24), (47), (59), (431) and (702) for the B-class, and asteroids
(213), (225), (302), (419), (704) and (762) for the F-class
\citep[taxonomy classification from NASA \emph{Planetary Data System}
  at {\tt http://pds.jpl.nasa.gov/}, and data from][]{RGH11, RGH12,
  Assandrietal2012, Giletal14, Beletal17}. Our
sample of B-class asteroids was collected using objects that were
classified as B both in SMASS and in the Tholen system; the
sample of F-class asteroids was built using only 'pure' F-objects in
the Tholen's classification, that is, without considering objects
classified as FC, CF, or similar ambiguous cases. Data points for
B-class and F-class asteroids, together with their best-fits, are
shown in Fig.~\ref{Fig_PQ}.  
  Best-fit parameters for all these curves are given in Table~\ref{Tab_Best-Fit}.  

The inversion angle (constrained by just one single point in the
negative polarization branch) is between the value of $\sim
16\,\degr$\ and $\sim 19\degr$, which are typical for the F-class and B-class,
respectively. We note, however, that the \ainv\ values of two
F-class asteroids determined by \citet{Beletal17} are actually $\ga
17\degr$, fully consistent with the value for Bennu. Based on
its \ainv\ value, Bennu could be either a B-class or an F-class
asteroid.

Due to the shortage of data points in the negative branch, the slope
at the inversion angle, ${\rm d}\pq/{\rm d}\lambda (\alpha=\ainv)$, and (even more) the
\pmin parameters are not well constrained. 
The slope, however, is closer to the values displayed by F-class asteroids. 

Taking at face value our best-fit solution, 
we can also in principle derive an estimate of the geometric albedo of Bennu.
Since \pmin is poorly determined and the trend in the positive polarisation
branch seems to be not very linear, and might lead us to use an exceedingly 
steep estimate, we prefer to use here the relation proposed
by \citet{Celetal15} between the albedo and the so-called $\Psi$
parameter, defined as $\Psi = \pq(30\degr) - \pq(10\degr)$. We found
for Bennu $\Psi = 6.083 \pm 0.266$ and a corresponding formal solution for the albedo of
$0.059 \pm 0.003$.  This value, to be confirmed when a better
sampling of the phase-polarisation curve is available, 
fits nicely the value found for the
F-class by \citet{Beletal17}, namely $0.058 \pm 0.011$. The same authors found
for the B-class an average albedo of $0.083\pm0.034$.
By using the $\Psi$ parameter and the
data shown in Fig.~\ref{Fig_PQ}, we found for the B and F classes
the values of $0.085 \pm 0.004$ and $0.057 \pm 0.014$, respectively.

Four of our seven data points for Bennu were taken at phase-angle 
$>40\degr$. Our best-fits for B and F asteroids may be
extrapolated at larger phase-angles, and Fig.~\ref{Fig_PQ} clearly
shows that Bennu's data are much more consistent with what predicted
for F-class rather than B-class asteroids.

\section{Discussion and Conclusions}\label{Sect_Discussion}
Until recently, a meaningful comparison with data for other asteroids
at large phase-angles would not have been possible, because only a few
moderate- or high-albedo near-Earth asteroids, supposed to have a
completely different thermal history and composition than Bennu, had
been observed. However, very recently \citet{Itetal18} and
\citet{Devetal18} carried out polarimetric observations of the NEO
(3200) Phaethon at very large phase-angles.  The analysis of their
data led \citet{Devetal18} to conclude that Phaethon is most likely a
collisional fragment of the B-class asteroid (2) Pallas, in particular
a fugitive from Pallas' family, confirming previous results based
exclusively on spectral reflectance data by \citet{deLetal10}. The
observations by \citet{Devetal18} and \citet{Itetal18} are also shown in
Fig.~\ref{Fig_PQ}. We note that at phase-angles $\sim 100\degr$ (not
shown in Fig.~\ref{Fig_PQ}), the two polarimetric datasets are not
consistent: around phase-angle 105\degr, polarisation values from
\citet{Itetal18} reach about 50\,\% while at phase-angle
101\degr\ \citet{Devetal18} measure only 38\,\%.  An explanation for
this discrepancy, possibly due to surface heterogeneity of Phaethon,
is beyond the scope of this paper. However, at phase-angles 
$\la 60\degr$, it clearly appears that Bennu exhibits values of linear
polarization much higher than those of Phaethon. Our measurements, as
well as the very recent observations of (152679) 1998 KU2 (Kuroda et
al. 2018) are the highest ever found for a NEO observed at large phase
angles.  Another solar system object that exhibit such high
polarisation values is comet C/1995 O1 (Hale-Bopp), which was observed by
\citet{ManBas00} in a 9\,nm wide gas-free filter centred around
$\lambda=694$\,nm. Hale-Bopp measurements are also shown in
Fig.~\ref{Fig_PQ}. The resemblance between the polarimetric data of
Bennu and those of Hale-Bopp is striking, and although reinforces the
conclusion about similarity between F-type objects and cometary dust
proposed by \citet{KolJok97}, it should be treated with caution, due
to the big differences between the two physical environments: in one
case we have an obviously active and very large comet, in the other
one a much smaller asteroid with no detection of a gas or dust coma.

Our data show that the phase - polarization curve of Bennu is steeper
than the extrapolation of phase-polarisation curves of main-belt
B-class asteroids observed at phase angles $\la 30\degr$, and
resembles that of F-class asteroids. At phase-angles $\la 30\degr$,
the polarimetric behaviour of Bennu is probably in between that
typical of the F and B classes. 
Dynamical modelling has suggested
that Bennu has most likely have evolved from the Inner part of the Main
Belt \citep[IMB; $a<2.5$\,au,][and references therein]{Waletal13}
possibly within either the ``New'' Polana or Eulalia families
\citep{Botetal15}. 
If this is true, taking into account the finding by \citet{Celetal01} that most
of Polana family's members belong to the F-class, our polarimetry
might support the hypothesis that Bennu originated in that
family, since Eulalia is a C-class asteroid.

Based on what is currently known about the similarity in polarimetric
behaviour of the asteroid F-class and of some cometary bodies,
(101955) Bennu 
could be compatible with cometary characteristics and a possible cometary
origin. According to \citet{deMeoBin08}, the estimated
fraction of comets likely present in the NEO population should be
$\,8\% \pm 5\,\%$. In a search for such objects, NEOs exhibiting the
typical behaviour of the F-class should be considered as highest
priority candidates, because, whatever can be the physical 
or evolutionary mechanism causing
an anomalously low value of the polarisation inversion angle, 
this rare property is shared by at least a few bodies 
belonging to the cometary population and by others belonging to the asteroid 
F taxonomic class.

We note that the conventional separation between
asteroids and comets has become increasingly uncertain in recent
years, also after the discovery of the so-called main belt comets
\citep{HsiehJewitt}. It is possible that the target of OSIRIS-REx
is an object borderline between typical asteroids and typical comets
\citep{Novetal12,Novetal14}.  A possible link with cometary objects
would have obvious consequences about the interpretation of the future
in situ exploration of Bennu by OSIRIS-REx and of the sample of
material brought back to the Earth. Moreover, we note that Bennu is
also one of the few objects for which an accurate measurement of a
Yarkovsky-driven drift in orbital semi-major axis has been obtained
\citep{Faretal13}, and used also to calibrate the most recently
developed methods of computation of the ages of asteroid dynamical
families \citep{Miletal14,Miletal17}. In this respect, it is clear that
the possibility that some (even weak) cometary activity could affect
the measurements of a Yarkovsky-driven evolution must be carefully
assessed.

\section*{Acknowledgments}
This work is based on observations collected at the European
Organisation for Astronomical Research in the Southern Hemisphere
under ESO programmes 0100.C-0771 and 2101.C-5011.

\label{lastpage}

\end{document}